\documentclass[aps,pre,preprint,groupedaddress]{revtex4-1}
\usepackage{amsmath,amssymb}  
\usepackage{amsthm}
\usepackage{graphicx} 
\usepackage{subfigure}
\newtheorem{theorem}{Theorem}[section]

\theoremstyle{remark}
\newtheorem{exa}[theorem]{Example}
\newtheorem{rem}[theorem]{Remark}

\newcommand{\ceil}[1]{\lceil{#1}\rceil}

\begin{document}
\title{Space-Time Duality and High-Order Fractional Diffusion}
\author{James F. Kelly}
\author{Mark M. Meerschaert}
\affiliation{Department of Statistics and Probability, Michigan State University, East Lansing, MI 48824}
\date{\today}
\begin{abstract}
Super-diffusion, characterized by a spreading rate $t^{1/\alpha}$ of the probability density function $p(x,t) = t^{-1/\alpha} p \left( t^{-1/\alpha} x , 1 \right)$, where $t$ is time, may be modeled by space-fractional diffusion equations with order $1 < \alpha < 2$.  Some applications in biophysics (calcium spark diffusion), image processing, and computational fluid dynamics utilize integer-order and fractional-order exponents beyond than this range ($\alpha > 2$), known as high-order diffusion, or hyperdiffusion.  Recently, space-time duality, motivated by Zolotarev's duality law for stable densities, established a link between time-fractional and space-fractional diffusion for $1 < \alpha \leq 2$.  This paper extends space-time duality to fractional exponents $1<\alpha  \leq 3$, and several applications are presented.  In particular, it will be shown that space-fractional diffusion equations with order $2<\alpha  \leq 3$ model sub-diffusion and have a stochastic interpretation.  A space-time duality for tempered fractional equations, which models transient anomalous diffusion, is also developed.
\end{abstract}
\pacs{2.30.Uu, 5.10.Gg, 5.40.-a, }
\maketitle
\section{Introduction}
Non-Fickian, or anomalous, diffusion is observed in many areas of physics, including hydrology \cite{deng2006parameter, phanikumar2007separating, haggerty2002power}, turbulent transport \cite{del2006fractional}, and biophysics \cite{fedotov2008non, jeon2011vivo}.  Anomalous \emph{super-diffusion} is characterized by a spreading rate $t^{1/\alpha}$ of the probability density function $p(x,t) = t^{-1/\alpha} p \left( t^{-1/\alpha} x , 1 \right)$ that is faster than the classical $t^{1/2}$ rate predicted by Fickian diffusion \cite{metzler2000random}, where $t$ is time, while anomalous \emph{sub-diffusion} is characterized by a spreading rate that is slower than $t^{1/2}$.  Fractional PDEs (FPDEs), where local time- and space-derivatives are replaced by non-local fractional derivatives, are often used to study anomalous diffusion.  FPDEs with a $\gamma$-fractional derivative in time and an $\alpha$-fractional derivative in space lead to a scaling rate of $t^{\gamma/\alpha}$.  Sub-diffusion may be modeled by a time-fractional derivative (e.g., Caputo derivative) with order $\gamma<1$ and a second derivative in space ($\alpha =2$) \cite{metzler2000random}, whereas super-diffusion may be modeled by a space-fractional derivative (e.g., Riemann-Liouville derivative) of order $1<\alpha <2$ and a first order derivative in time ($\gamma =1$) \cite{metzler2004restaurant}.  These FPDEs may be derived from a continuous time random walk (CTRW) framework: time-fractional diffusion equations involve long-waiting times between particle jumps, where the chance of waiting longer than some time $t>0$ is proportional to $t^{-\gamma}$, while space-fractional diffusion equations involve long particle jumps, where the chance of jumping longer than some distance $x>0$ is proportional to $x^{-\alpha}$.  Transient anomalous sub- and super-diffusion, which transition from early-time anomalous behavior to late-time diffusive behavior, may be modeled with tempered time-fractional \cite{meerschaert2008tempered} and space-fractional \cite{baeumer2010tempered, cartea2007} derivatives, respectively.

Recently, we have established a link between time-fractional and space-fractional diffusion equations, called space-time duality \cite{baeumer2009space, kellyduality}.  Zolotarev  \cite{zolotarev1961expression, zolotarev1986one} first proved a duality law between stable densities with indices $1<\alpha \leq 2$ and $1/2 \leq 1/\alpha < 1$.  The duality principle was applied to the space-fractional diffusion equation in \cite{baeumer2009space}, and later to the space-fractional advection-dispersion equation in \cite{kellyduality}.  The latter study was motivated by a controversy in river-flow hydrology: both space-fractional dispersion (diffusion) equations and time-fractional PDEs provide reasonably good fits to breakthrough curve (BTC) measurements \cite{kelly_fracfit}.  From a stochastic point of view, space-time duality established a connection between long, power-time waiting times and long negative jumps, thereby justifying a space-fractional PDE for modeling retention of contaminant particles.  In short, a particle that rests while the plume moves downstream ends up in the same position as a particle that moves downstream, but then makes a long upstream jump.

In both \cite{baeumer2009space} and \cite{kellyduality}, the equivalence was restricted to space-fractional PDEs modeling super-diffusion $(1 < \alpha < 2)$.  The equivalent time-fractional equation has order $\gamma = 1/\alpha$.  Space-fractional derivatives of order $\alpha > 2$ have recently been used to model sub-diffusion of calcium sparks in cardiac myocytes by Chen et al. \cite{chen2013} and Tan et al. \cite{tan2007anomalous}, exhibiting good agreement with experimental data.  This sub-diffusion results from the multi-scale nature of cytoplasm, which has polymer networks and complex macro-molecules that immobilize diffusing particles.  Recall that time-fractional PDEs are often used to model sub-diffusion since the time-fractional Caputo derivative results from long waiting times in the CTRW formalism.   A question arises: can the space-fractional model with order  $2<\alpha \leq 3$ proposed in \cite{tan2007anomalous, chen2013} be linked with time-fractional  \cite{fedotov2008non} and CTRW \cite{jeon2011vivo} diffusion models also used in biophysics?  Space-fractional exponents with $\alpha >2$ (high-order diffusion, or hyperdiffusion) are also found in fluid mechanics  \cite{frisch2008hyperviscosity}, image processing \cite{wei1999generalized}, and transport of cosmic rays \cite{tawfik2018}.

The goal of this paper is to extend space-time duality to fractional (and integer) spatial derivatives of order $1<\alpha  \leq 3$.  Our duality result shows how both super-diffusion and sub-diffusion can be modeled by a space-fractional PDE.  Then we illustrate the method with applications to the time-fractional diffusion wave equation, multi-dimensional time-changed Brownian motion, and tempered fractional diffusion.  In Section II, we briefly review the space-fractional diffusion equation and hyperdiffusion.   Section III generalizes the space-time duality argument presented in \cite{kellyduality} to all space-fractional exponents $1 <\alpha \leq 3$.  Section IV connects solutions of the time-fractional diffusion-wave equation to a corresponding system of space-fractional diffusion equations.  A governing equation for subordinated multi-dimensional Brownian motion is proposed in Section V using a  vector space-fractional PDE.  Section VI extends space-time duality to tempered fractional diffusion, followed by conclusions in Section VII. 
\section{Space-Fractional Diffusion}
The two-sided space-fractional diffusion equation is given by \cite[Equation (1.26)]{meerschaert2012}
\begin{equation}
\frac{\partial}{\partial t} u(x,t) =  \left(\frac{1 + \theta}{2}\right) C \frac{\partial^{\alpha}}{\partial x^{\alpha}} u(x,t)+  \left(\frac{1 - \theta}{2}\right) C \frac{\partial^{\alpha}}{\partial (-x)^{\alpha}} u(x,t)
\label{fde}
\end{equation}
where $C$ is a fractional diffusion coefficient,  the fractional index is $\alpha >1$, and the skewness is $\theta \in[-1,1]$.  The positive (left) and negative (right) Riemann-Liouville (RL) fractional derivatives are defined by \cite[p. 87]{kilbas2006theory}
\begin{subequations}
\label{rlderivs}
\begin{equation}
\frac{\partial^{\alpha}}{\partial x^{\alpha}} f(x) = \frac{1}{\Gamma(n - \alpha)} \frac{\partial^n}{\partial x^n} \int_{-\infty}^{x}  f(y) (x-y)^{n -1 - \alpha} \, dy
\label{posrl}
\end{equation}
\begin{equation}
\frac{\partial^{\alpha}}{\partial (-x)^{\alpha}} f(x) = \frac{(-1)^n}{\Gamma(n - \alpha)} \frac{\partial^n}{\partial x^n} \int_x^{\infty}  f(y) (y-x)^{n-1 - \alpha} \, dy ,
\label{negrl}
\end{equation}
\end{subequations} 
where $n = \ceil{\alpha}$ and $\Gamma(z)$ is the Gamma function.  For $1< \alpha \leq 2$ subject to an impulse initial condition $u(x,0) = \delta(x)$, the fundamental solution of \eqref{fde} is a stable probability density function (PDF) with index $\alpha$ and skewness $\theta$ \cite{mainardi2001fundamental}.  In river-flow hydrology, breakthrough curve measurements of relative concentration $u(x,t)$ with $x$ fixed are well fit by negatively-skewed ($\theta = -1$) PDFs \cite{kelly_fracfit}.  
%
%

For the special case of $\theta = -1$, \eqref{fde} reduces to the negatively skewed space-fractional diffusion equation 
\begin{equation}
\frac{\partial}{\partial t} u(x,t) = C \frac{\partial^{\alpha}}{\partial (-x)^\alpha} u(x,t).
\label{fde2}
\end{equation}
The coefficient $C$ is chosen such that the eigenvalues of \eqref{fde2} have a non-positive real part so energy is not created.  Denote the Fourier transform (FT) of $u(x,t)$ by $\hat{u}(k,t)$ and apply a FT to \eqref{fde2}, yielding
\begin{equation*} 
\frac{\partial}{\partial t} \hat{u}(k,t) = C (-ik)^{\alpha} \hat{u}(k,t) .
\label{fde2hat}
\end{equation*}
Since the real part of $C (-ik)^{\alpha}$ is $C \cos ( \pi \alpha /2)$, we take $C = (-1)^{m+1}$ where $2m - 1 < \alpha < 2m + 1$ and $m \in \mathbb{N}$ to produce eigenvalues with non-positive real part.  In particular, $C=1$  if $1 < \alpha \leq 3$ and $C=-1$ if $3 < \alpha \leq 5$.  Under this condition, \eqref{fde2} reduces to a \emph{hyperdiffusion equation} \cite{frisch2008hyperviscosity, baeumer2015higher}
\begin{equation}
\frac{\partial}{\partial t} u(x,t) =  (-1)^{m+1} \frac{\partial^{\alpha}}{\partial (-x)^{\alpha}} u(x,t) .
\label{hyperdiff}
\end{equation}  
For integer $\alpha = 2m$, \eqref{hyperdiff} is used in turbulence modeling \cite{frisch2008hyperviscosity}, stabilizing numerical methods such as the spectral element method \cite{ullrich2018impact}, and modeling the transport of cosmic rays \cite{malkov2015cosmic}.  In the remainder of this paper, we consider \eqref{fde2} with $C= 1$ for $1 < \alpha \leq 3$, which is a special case of \eqref{hyperdiff}.

We consider solutions with an impulse initial condition $u(x,0) = \delta(x)$.  For $1 < \alpha \leq 2$, solutions to \eqref{fde2} are negatively skewed stable densities \cite{mainardi2001fundamental}, which model anomalous diffusion where particles experience large jumps in the negative direction.  This equation, complemented with a drift term, successfully models contaminant transport in rivers \cite{phanikumar2007separating, chakraborty2009parameter}, as well as source identification problems in groundwater hydrology \cite{zhang2016backward}, where $u(x,t)$ is the release location/time PDF.  These hydrology applications assume a fractional exponent $1 < \alpha \leq 2$, so that the contaminant particles experience super-diffusion and there is stochastic interpretation to $u(x,t)$.  
\begin{rem}
The term ``hyperdiffusion'' has several usages in the literature. For example, Metzler et al. \cite{metzler2012} define hyperdiffusion as a process with mean-squared displacement that has a scaling rate of $t^{\alpha}$, where $\alpha > 2$.  In this paper, the term ``hyperdiffusion'' refers to  the FPDE \eqref{hyperdiff} with $\alpha > 2$ and its solutions.  Hyperdiffusion (or hyperviscosity) is popular in turbulence modeling and computational fluid dynamics (CFD), where integer powers greater than two are used to stabilize numerical methods by reducing the range of scales over which dissipation acts \cite{frisch2008hyperviscosity}.  Hyperdiffusion is used in spectral element models to damp high-order modes and eliminate numerical noise \cite{ullrich2018impact}.  The most commonly used value for hyperdiffusion is $\alpha = 4$ ($m=2$) \cite{satoh2008, ullrich2018impact}.  Wei \cite{wei1999generalized} applied integer-order hyperdiffusion for image denoising and edge detection problems, while Malkov and Sagdeev \cite{malkov2015cosmic} derived a hyperdiffusion model with $\alpha = 4$ ($m=2$) for cosmic ray transport.  Fractional-order hyperdiffusion with orders larger than two have also been used in the surface generation of proteins by Hu et al. \cite{hu2013high} and modeling calcium sparks in cardiac myocytes by Tan et al. \cite{tan2007anomalous}.  Recently, Tawfik et al. \cite{tawfik2018} used a space-time hyperdiffusion equation with a Riesz derivative in space of order $\alpha > 2$ and Caputo derivative in time of order $0 < \gamma < 1$ to model cosmic rays. 
\end{rem}
\section{Space-Time Duality}
Although space-time duality was first noted using stable PDFs \cite{zolotarev1961expression, zolotarev1986one}, the basic idea may be illustrated using Fourier transforms and dispersion relationships.   Applying a space-time FT to \eqref{fde2} using the relationship $\int_{-\infty}^{\infty} \frac{\partial^{\alpha}}{\partial (-x)^{\alpha}} f(x) e^{-ikx} \, dx = (-ik)^{\alpha} \hat{f} (k)$ yields a dispersion relationship $i \omega = (-ik)^{\alpha}$, where $\omega$ is angular frequency and $k$ the wavenumber, and $\hat{f}(k)$ is the spatial FT of $f(x)$.  Formally take the $\alpha$-th root, yielding an equivalent dispersion relationship $(i \omega)^{\gamma} = -ik$, where $\gamma = 1/\alpha$, which characterizes a \emph{time-fractional} PDE of order $\gamma < 1$.  

Although this argument is heuristic, it motivates a Fourier-Laplace transform (FLT) argument first presented in \cite{kellyduality}.  In \cite{kellyduality}, we restricted our attention to fractional orders $1 < \alpha \leq 2$ in \eqref{fde2} with $C = 1$.  In this section, this restriction on $\alpha$ is relaxed, allowing the fractional order to be larger than two and less than or equal to three and providing a stochastic model for hyperdiffusion.  Our motivation comes from Hu et al. \cite{hu2013high}: ``Currently, most attention in the field is paid to the fractional derivatives of order less than 2. High-order fractional derivatives are hardly used, partly due to the limited understanding of their physical meanings."  In this section, we assign a physical meaning to \eqref{fde2} with $2 < \alpha \leq 3$ using a space-time duality argument.

Define the FLT of $u(x,t)$ via
\begin{equation}
\overline{u}(k,s) = \int_0^{\infty} \int_{-\infty}^{\infty} u(x,t) e^{-s t} e^{-i k x} \, dx \, dt
\label{flt}
\end{equation}
and the Laplace transform by $\tilde{u} (x,s)$.  Then apply \eqref{flt} to \eqref{fde2} with $C=1$, yielding
\begin{equation}
\overline{u} (k,s) = \frac{1}{s - (-ik)^{\alpha}} .
\label{flt1}
\end{equation}
The inverse FT of \eqref{flt1} can be expressed as \cite[(4.8.18)]{morse1953methods}
\begin{equation}
\tilde{u} (x, s) = \frac{1}{2 \pi} \lim_{R\to\infty} \int_{-R + i\tau}^{R + i\tau}\frac{e^{ikx}}{s - (-ik)^{\alpha}} \, dk ,
\label{invfourier}
\end{equation}
where $\tau>0$ is chosen to avoid the branch cut along the negative real axis.  

For $1 < \alpha \leq 3$,  the integrand of \eqref{invfourier} has a single, simple pole at $k^{*}= is^{1/\alpha}$ and remains analytic for all other points in the upper half-plane (UHP) for any choice of $1<\alpha \leq 3$.  To prove this, write the wavenumber in polar form $k = |k| e^{i \theta}$, where $|\theta| \leq \pi$ is the phase angle.  The poles $k^{*}$ then satisfy
\begin{equation}
|k^{*}|^{\alpha} e^{i \alpha ( \theta - \pi /2)} = s
\label{poleeqn}
\end{equation}
where $s$ is positive and real.  Hence, the phase angle satisfies $\alpha (\theta - \pi /2) = 2 \pi n$ with $n \in \mathbb{N}$.  Since we are only interested in poles that reside in the UHP,  take $0 < \theta < \pi$.  Solving for $n$ yields $-\alpha / 3 < n < \alpha /3$. Hence, if $1<\alpha \leq 3$, the only integer solution is $n=0$, implying that only one pole lies in the UHP.  If $3<\alpha  \leq 5$, then the coefficient on the right hand side of \eqref{hyperdiff} is negative, yielding a FLT of $\overline{u}(k,s) = \left(s + (-ik)^{\alpha}\right)^{-1}$.  Repeating the pole calculation yields at least \emph{two} poles in the UHP for $3 < \alpha \leq 5$, while for $\alpha > 5$, there are at least \emph{three} poles in the UHP.  Hence, the complex plane argument described below is not applicable and we cannot assign a stochastic interpretation to the space-fractional diffusion equation for $\alpha > 3$.

By converting the path of integration in \eqref{invfourier} into a closed contour in the upper half-plane by attaching a semi-circle of radius $R$ (see Appendix A in \cite{kellyduality} for details), \eqref{invfourier} is evaluated using the Cauchy residue theorem as
\begin{equation}
\tilde{u} (x, s) = \gamma s^{\gamma -1} \exp \left( -x s^{\gamma} \right) 
\label{flt2}
\end{equation}
where $1/3 \leq \gamma = 1 / \alpha < 1$.  The contribution along the semi-circle $C_R$ vanishes as $R \rightarrow \infty$ using the bounds in Appendix A of \cite{kellyduality}.

Inverting the LT yields
\begin{equation}
u(x,t) = \gamma h_{\gamma}(x,t)
\label{usol1}
\end{equation}
where $h_{\gamma}(x,t)$ is the inverse stable density (see Remark \ref{invstblrem} below) with index $\gamma$ \cite{meerschaert2013inverse}.  To derive the governing equation of the inverse stable density, take the FT of \eqref{flt2}, yielding
\begin{equation}
\tilde{u} (x, s) = \frac{ \gamma s^{\gamma -1}}{ik + s^{\gamma}} .
\label{flt3}
\end{equation}
Recall that the LT of the Caputo derivative $\partial^{\gamma}_t u(x,t)$ is given by $\mathcal{L}_t \left[ \partial^{\gamma}_t u(x,t) \right] = s^{\gamma} \tilde{u} (x,s) - s^{\gamma -1} u(x,0)$ for $0 < \gamma < 1$ \cite[Equation (1.27)]{mainardi2010fractional}.  Cross-multiply and invert, yielding
\begin{equation}
\partial^{\gamma}_t u(x,t) = -\frac{\partial}{\partial x} u(x,t); \quad u(x,0) = \gamma \delta(x) ,
\label{disp2xt}
\end{equation}
which is valid for any $1/3 \leq \gamma < 1$.  Hence, we have transformed the space-fractional equation \eqref{fde2} into an equivalent time-fractional equation \eqref{disp2xt} on the half-axis.  This result extends the results of \cite{baeumer2009space} and \cite{kellyduality} to a larger range of fractional (and integer) exponents $1<\alpha \leq 3$ and time-fractional exponents $1/3 \leq \gamma<1$.  

For $\alpha \neq 2,3$, the spatial nonlocality of the negative RL derivative is exchanged for the temporal nonlocaity of the Caputo derivative.  The time-fractional equation \eqref{disp2xt} governs the long term limit of a random walk where the particles experience power-law waiting times $T$ with tail probability $P(T > t) \approx t^{-\gamma}$ for $t \gg 1$.  Hence, we can assign a stochastic intepretation to \eqref{fde2} for $2<\alpha  \leq 3$: the fractional order $\alpha$ codes long, power-law waiting times that scale like $t^{-1/\alpha}$.  Note that the tail of the waiting time distribtion associated with \eqref{fde2} is \emph{heavier} than those considered in \cite{kellyduality}, indicating a higher probability of very long waiting times. 
\begin{rem}
\label{invstblrem}
The time-fractional equation \eqref{disp2xt} is the governing equation of the inverse stable subordinator \cite{meerschaert2013inverse}
\begin{equation} 
E_t = \mbox{inf} \left\{ x>0 : D_x > t \right\}
\label{invsub}
\end{equation}
that models the first passage times of the stable subordinator $t= D_x$, where $D_x$ has density $g(t,x)$ with Laplace transform $e^{-xs^{\gamma}}$.  From a CTRW perspective, the inverse process $E_t$ models the local times of particles undergoing long waiting times.
\end{rem}
\begin{exa}
The inverse $\gamma$-stable subordinator of order $\gamma = 1/3$ satisfies the integer-order PDE
\begin{equation}
\frac{\partial}{\partial t} h_{1/3}(x,t) = -\frac{\partial^{3}}{\partial x^3} h_{1/3}(x,t)
\label{orderonethird}
\end{equation}
which is a linearized KdV equation \cite{whitman} used to model long wavelength water waves.  Equation \ref{orderonethird} may be evaluated in closed form \cite[Equation (2.10.3)]{zolotarev1986one}
\begin{align}
h_{1/3}(x,t) =& \frac{1}{\gamma} \mathcal{F}_x^{-1} \left[ \exp \left( t(-ik)^3 \right) \right] \nonumber \\
         =& \frac{3}{2 \pi} \int_{-\infty}^{\infty} \exp \left(i (kx + tk^3) \right) \, dk \nonumber \\
         =& \frac{3}{(3t)^{1/3}} \mbox{Ai} \left( \frac{x}{(3t)^{1/3}} \right)
\label{airysol}				
\end{align}
where $\mbox{Ai} (z)$ is the Airy function.   Hence, $h_{1/3}(x,t)$ spreads at rate $t^{1/3}$, which is clearly sub-diffusive.  
\end{exa}
\begin{rem}
The density $h_{\gamma}(x,t)$ is self-similar with a scaling relationship $h_{\gamma}(x,t) = t^{-1/\alpha} h_{\gamma}(x t^{-1/\alpha},1)$ \cite{meerschaert2013inverse}.  We can distinguish three types of behavior: (i) if $1<\alpha < 2$, the plume spreads faster than the diffusive rate of $t^{1/2}$;
 (ii) if $\alpha =2$, the solution is classically diffusive; and (iii) if $2<\alpha \leq  3$, the solution spreads slower than the diffusive rate of $t^{1/2}$.  Hence, a wide range of anomalous diffusion may be modeled with a negatively skewed space-fractional diffusion equation. 
\end{rem}
\begin{rem}
\label{positiveskew}
Space-time duality may be applied to the positively-skewed case $\theta = 1$ on the \emph{negative} half-axis $x<0$ by the same argument.  Zolotarev wrote a general duality law involving stable PDFs for $\alpha \leq 2$ \cite[Equation 2.3.3]{zolotarev1986one} and trans-stable distributions for $\alpha > 2$ \cite[Equation 2.11.7]{zolotarev1986one}.  This duality law is valid for a range of skewness parameters $\theta$.  These duality relations may be extended to the negative half-axis using the reflection property of stable and trans-stable PDFs.  Using these relationships, we derived a time-fractional equation involving both positive and negative temporal RL derivatives that is equivalent to \eqref{fde2} for $x<0$ in Appendix C of \cite{kellyduality}.  It should be possible to extend this result to the two-sided diffusion equation \eqref{fde} by a similar argument.  Unlike the negative spatial RL derivative, it is not known how to assign any physical meaning to a negative (right) temporal RL derivative, which models temporal nonlocality into the future.  
\end{rem}
\begin{rem}
It is also interesting to consider the physical meaning of a time derivative of order $\gamma>1$.  Some results in this direction can be found in \cite{baeumer2007} for the case $1<\gamma<2$.  For a diffusion with drift, introducing a fractional time derivative of order $1<\gamma<2$ results in a kind of superdiffusion, where the plume variance spreads like $t^{3-\gamma}$, see \cite[Section 6.2]{baeumer2007}.  We do not know whether there is a duality result for $\gamma>1$.
\end{rem}
%
%
\begin{rem}
\label{numerics}
Conservative explicit Euler \cite{baeumer2017boundary} and implicit Euler \cite{kellybcs2018} methods are available to solve \eqref{fde2} subject to the reflecting boundary condition \eqref{nofluxbc}.  Feng \cite{feng2018} proposed an unconditionally stable Crank-Nicolson scheme for fractional orders $2 < \alpha < 3$ that is first-order accurate in space and second-order accurate in time.  Baeumer et al. \cite[Proposition 4.2]{baeumer2015higher} proposed a stable scheme for \eqref{hyperdiff}  for any $\alpha$ that is high-order in space based on a Gr\"{u}nwald discretization \cite{meerschaert2006finite} with shift $m$, where $m$ is given by $2m -1 < \alpha < 2m + 1$.  
\end{rem}
\section{Time-Fractional Diffusion-Wave Equation}
A wide variety of anomalous phenomena can be modeled by the time-fractional diffusion-wave equation on the real line 
\begin{equation}
\partial_t^{\beta} u(x,t) = \frac{\partial^2}{\partial x^2} u(x,t) ,
\label{timefrac}
\end{equation}
where $0 < \beta \leq 2$, $\beta = 2 \gamma$, and the left hand side is the Caputo derivative of order $\beta$.  Equation \eqref{timefrac} interpolates between the diffusion equation ($\beta = 1$) and the wave equation ($\beta =2$).   For $0 < \beta <1$, \eqref{timefrac} models anomalous sub-diffusion and Hamiltonian chaos \cite{zaslavsky1994fractional}.  In particular, $u(x,t)$ is the limiting density of a CTRW with a Pareto (power-law) waiting time distribution $P(J_n > t) = Bt^{-\beta}$ \cite{baeumer2001stochastic}.  For $1 < \beta < 2$, \eqref{timefrac} models wave propagation in viscoelastic materials \cite{mainardi2010fractional}, including seismic waves \cite{mainardi1997seismic} and acoustic waves in biological media \cite{meerschaert2015stochastic}.       
\subsection{Analytical Solution}
Fundamental solutions to \eqref{timefrac} on the real line are computed using the initial condition $u(x,0) = \beta \delta(x)$.  For $1 < \beta \leq 2$, we impose the additional initial condition $u_t(x,0) = 0$.  The Laplace transform of the Caputo derivative with order $1 < \beta \leq 2$ is given by \cite[Equation (1.27)]{mainardi2010fractional}
\begin{equation}
\mathcal{L}_t \left[ \partial_t^{\beta} u(x,t) \right] = s^{\gamma} \tilde{u} (x,s) - s^{\gamma -1} u(x,0) - s^{\gamma -2} u_t(x,0)
\label{ltcaputo}
\end{equation}
while for $0 < \beta \leq 1$, the Laplace transform is merely the first two terms.  Apply a FLT to \eqref{timefrac}, yielding
\begin{equation}
\overline{u} (k,s) = \frac{\beta s^{\beta -1}}{k^2 + s^{\beta}} .
\label{flt11}
\end{equation}
Factor the denominator into $(s^{\beta/2} + ik)(s^{\beta/2} - ik)$ and expand in partial fractions,
yielding
\begin{equation}
\overline{u} (k,s) = \frac{\gamma s^{\gamma/2 -1}}{ik + s^{\gamma}} + \frac{\gamma s^{\gamma/2 -1}}{-ik + s^{\gamma}}
\label{flt22}
\end{equation}
where $1/2 < \gamma = \beta/2 \leq 1$.   Noting that the first term has a pole $k^{*} = i s^{\gamma}$ in the upper-half $k$ plane, and the second term has a pole $k^{*} =- i s^{\gamma}$ in the lower-half $k$ plane, we see that the first term has support on $x>0$ while the second term has support on $x<0$.  Applying an inverse FLT to each term in \eqref{flt22} yields a pair of \emph{one way fractional wave equations}
\begin{subequations}
\begin{equation}
\partial_t^{\gamma} u_{+}(x,t) = -\frac{\partial}{\partial x} u_{+}(x,t) \quad\text{for $x>0$ and }
\label{rightmoving}
\end{equation}
\begin{equation}
\partial_t^{\gamma} u_{-}(x,t) = \frac{\partial}{\partial x} u_{-}(x,t) \quad\text{for $x<0$.}
\label{leftmoving}
\end{equation}
\label{factoredeqns}
\end{subequations}
Much like the classical wave equation, \eqref{timefrac} consists of left and right moving components.  A similar decomposition was reported in \cite[Equation (5.4)]{meerschaert2015stochastic} for $1 \leq \beta \leq 2$.  The solution of \eqref{rightmoving} is the density of the inverse $\gamma$-stable subordinator 
\begin{equation}
h_{\gamma}(x,t) = \frac{t}{x^{1+ 1/\gamma}} g_{\gamma} \left( t x^{-1/\gamma} \right) ,
\label{invstablesub}
\end{equation}
where $g_{\gamma}(x)$ is the density of the  $\gamma$-stable subordinator with Laplace transform $e^{-s^{\gamma}}$.  The left-moving component is given by $u_{-}(x,t) = \gamma h_{\gamma}(-x,t)$.  Combining these two components yields
\begin{equation}
u(x,t) = \frac{\gamma}{2} h_{\gamma} (|x|,t) ,
\label{waveeqsol}
\end{equation}
which is also given in Mainardi et al. \cite[Equation (4.23)]{mainardi2001fundamental} using the Wright function.  Note that \eqref{waveeqsol} is continuous but not differentiable at $x=0$ with a ``cusp" at $x=0$ \cite[Proposition 6.1]{alrawashdeh2017applications}.  See also \cite{metzler2000random, carnaffan2017cusping}. 
\subsection{Duality Solution}
\label{waveeqdual}
\begin{figure}
\subfigure{\includegraphics[width=3in]{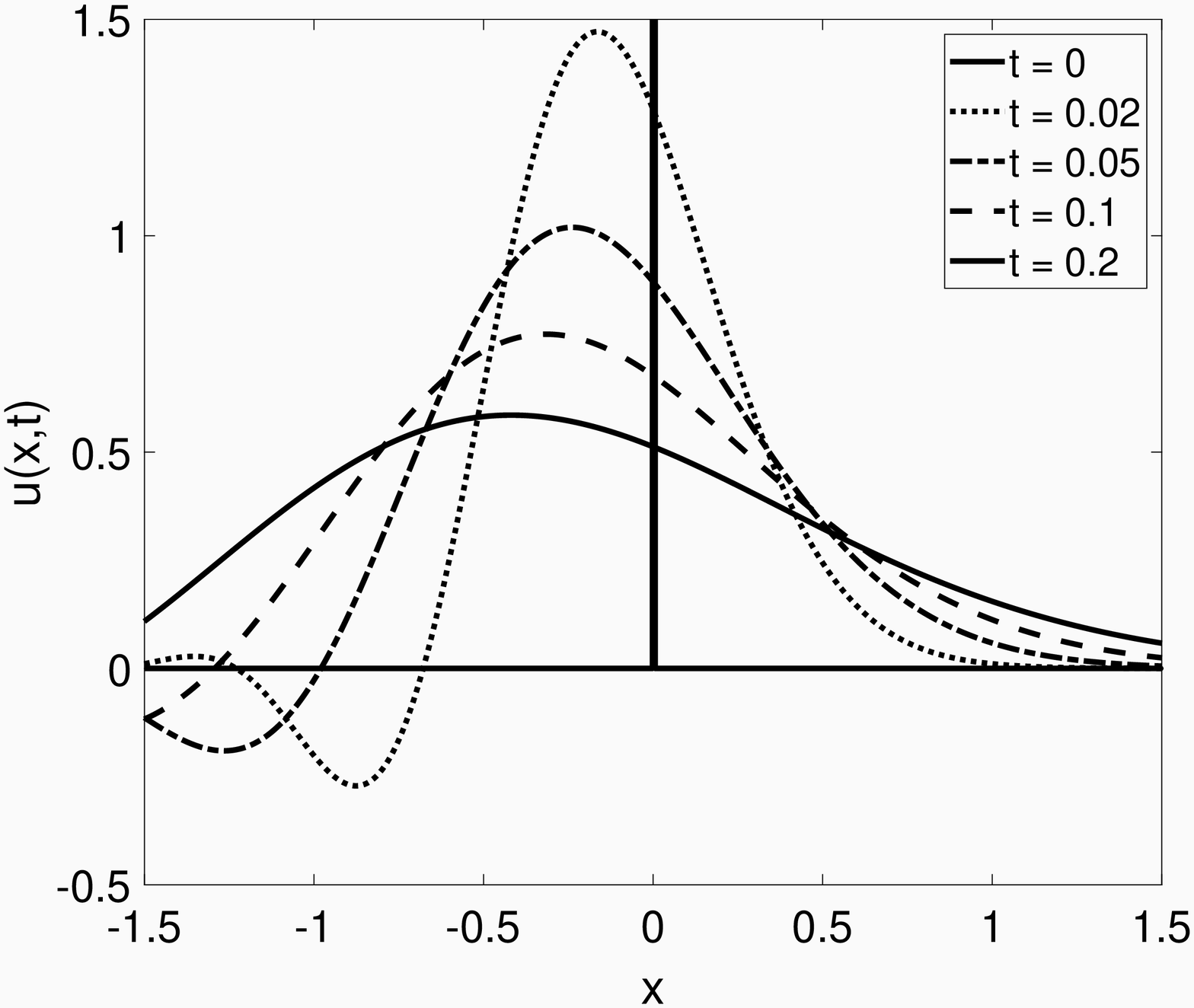}}
\hfil
\subfigure{\includegraphics[width=3in]{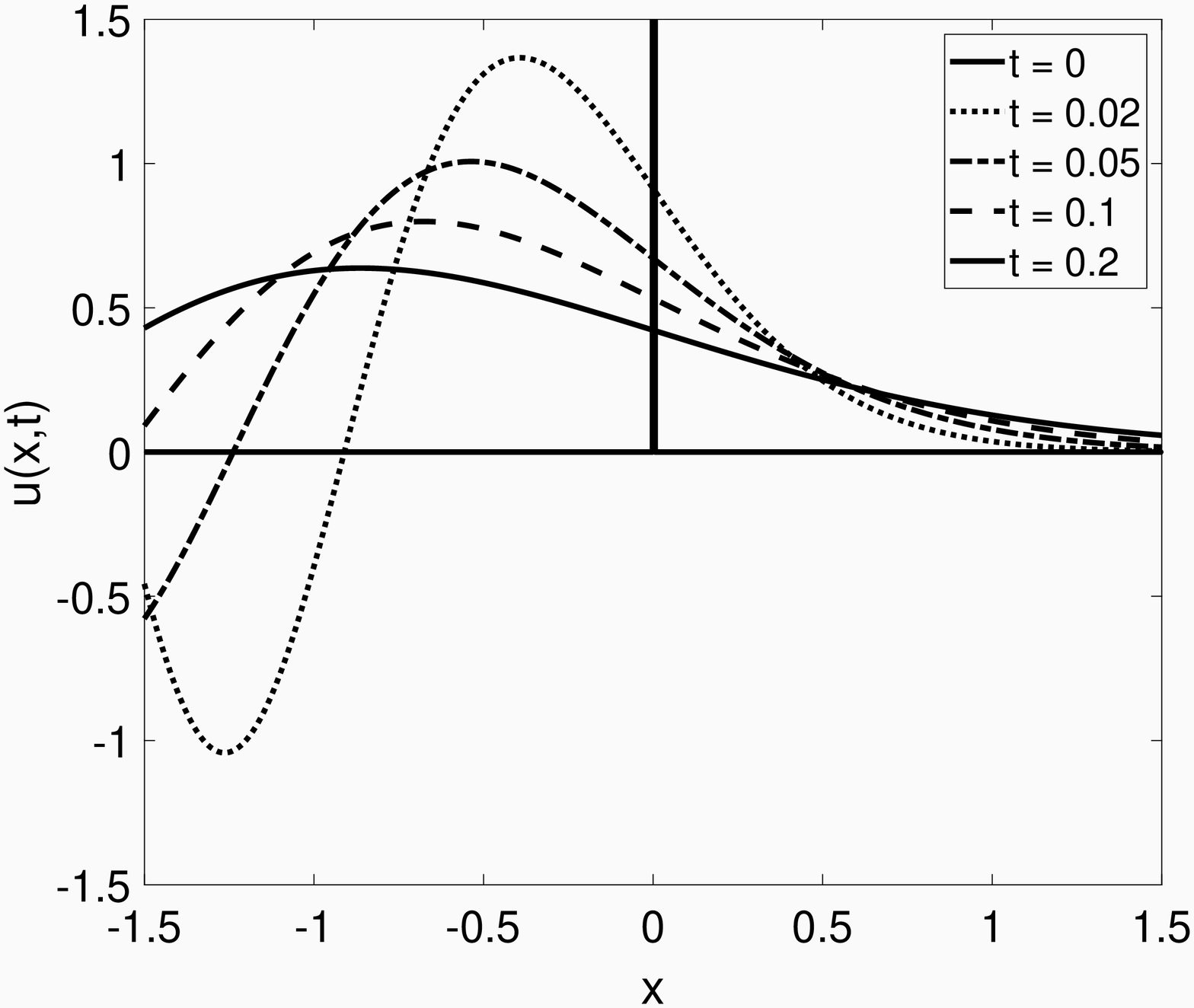}}
\caption{Numerical solutions of the space-fractional diffusion equation \eqref{fde2} on the entire real line using the implicit Euler scheme outlined in \cite{baeumer2017boundary} and \cite{kellybcs2018} for $\alpha = 2.5$ (left) and $\alpha = 3$ (right).  Note that the solutions are non-negative for $x>0$ but assume both positive and negative values for $x<0$.} 
\label{realaxisfig}
\end{figure}

By duality, the system of one way time-fractional equations \eqref{factoredeqns} may be converted into a system of space-fractional equations on the real line.
We see that $u_{+}(x,t)$ also solves \eqref{fde2} with $\alpha = 1/\gamma = 2/\beta$ and $C=1$.  Applying Remark \ref{positiveskew}, the solutions to \eqref{timefrac} also solve a system of space-fractional PDEs 
\begin{subequations}
\label{spacefrac2}
\begin{equation}
\frac{\partial}{\partial t} u(x,t) = \frac{\partial^{\alpha}}{\partial (-x)^{\alpha}} u(x,t) \quad\text{for $x>0$ and }
\label{spacefrac2a}
\end{equation}
\begin{equation}
\frac{\partial}{\partial t} u(x,t) = \frac{\partial^{\alpha}}{\partial x^{\alpha}} u(x,t) \quad\text{for $x<0$, }
\label{spacefrac2b}
\end{equation}
\end{subequations}
which may be expressed for any real $x$ via
\begin{equation}
\frac{\partial}{\partial t} u(x,t) = A_x^{\alpha} u(x,t)
\label{ucomp}
\end{equation}
using the operator
\begin{equation}
A^{\alpha}_x f(x) = \begin{cases}   
    \frac{\partial^{\alpha}}{\partial (-x)^{\alpha}} f(x) & x > 0 \\
    \frac{\partial^{\alpha}}{\partial x^{\alpha}} f(x) & x < 0, \\                                              
\end{cases}
\label{Aoperator}
\end{equation}
In the case of sub-diffusion ($2/3 \leq \beta < 1$), $2 <\alpha  \leq 3$, while for super-diffusion ($1 < \beta < 2$), $1 < \alpha  < 2$.
\begin{rem}
\label{stochasticinterp}
For $1 < \alpha \leq 2$, \eqref{spacefrac2a} complemented by the boundary condition \eqref{nofluxbc} govern spectrally negative L\'{e}vy motion conditioned to stay positive \cite{baeumer2016reflected}, while \eqref{spacefrac2b} governs spectrally positive L\'{e}vy motion conditioned to stay negative.  On the positive half-axis, particles may drift to the right or jump to the left.  On the negative half-axis, particles may drift left or jump to the right.
\end{rem}
\begin{rem}
\label{nopdf}
Note that solutions to either \eqref{spacefrac2a} or \eqref{spacefrac2b} on the \emph{entire} real line are not positive for $\alpha > 2$, which may be shown by calculating moments
using $\int_{-\infty}^{\infty} x^n u(x,t) \, dx = i^n \hat{u}^{(n)} (0,t)$.  Hence, these solutions on the real line are not PDFs.  Numerical solutions to \eqref{fde2} on the real line are shown in Figure \ref{realaxisfig} for $\alpha = 2.5$ and 3, illustrating this non-positivity.  
\end{rem}
\begin{rem}
\label{myo}
The space-fractional diffusion equation of order $\alpha = 2.25$ was proposed by Tan et al. \cite{tan2007anomalous} to model sub-diffusion of calcium sparks in the heart.  Since the space-fractional diffusion equation \eqref{fde2} of order $2<\alpha \leq 3$ is mathematically equivalent to a time-fractional diffusion equation of order $1/\alpha$, \eqref{fde2} is the limit of a CTRW with waiting times $J_n$ that are asymptotically Pareto with index $1/3 \leq \gamma < 1/2$.  Hence the space-fractional PDE with $2<\alpha \leq 3$ models anomalous sub-diffusion caused by particle sticking or trapping.
\end{rem}
\subsection{Numerical Experiments}
\begin{figure}
\subfigure{\includegraphics[width=3in]{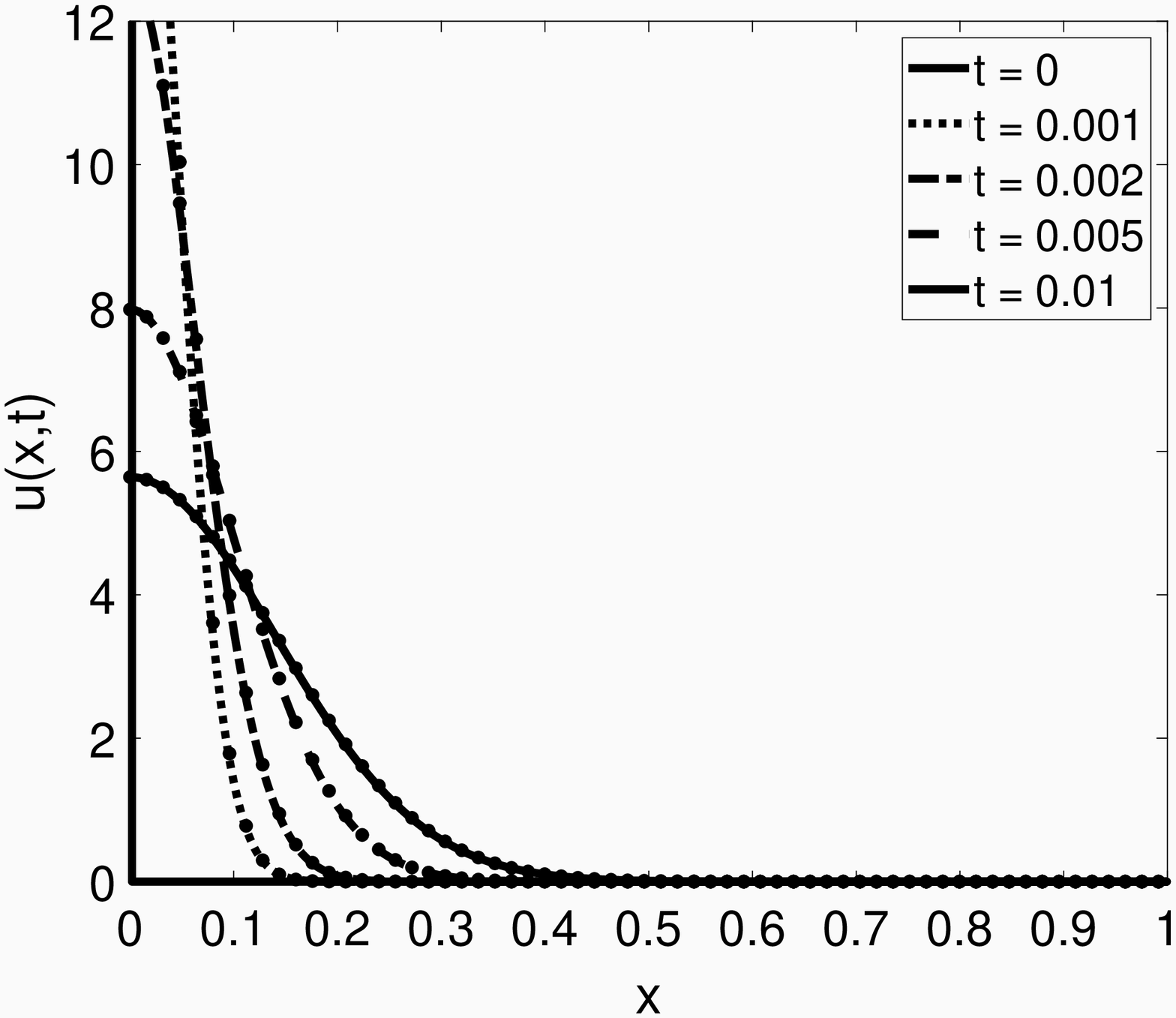}}
\hfil
\subfigure{\includegraphics[width=3in]{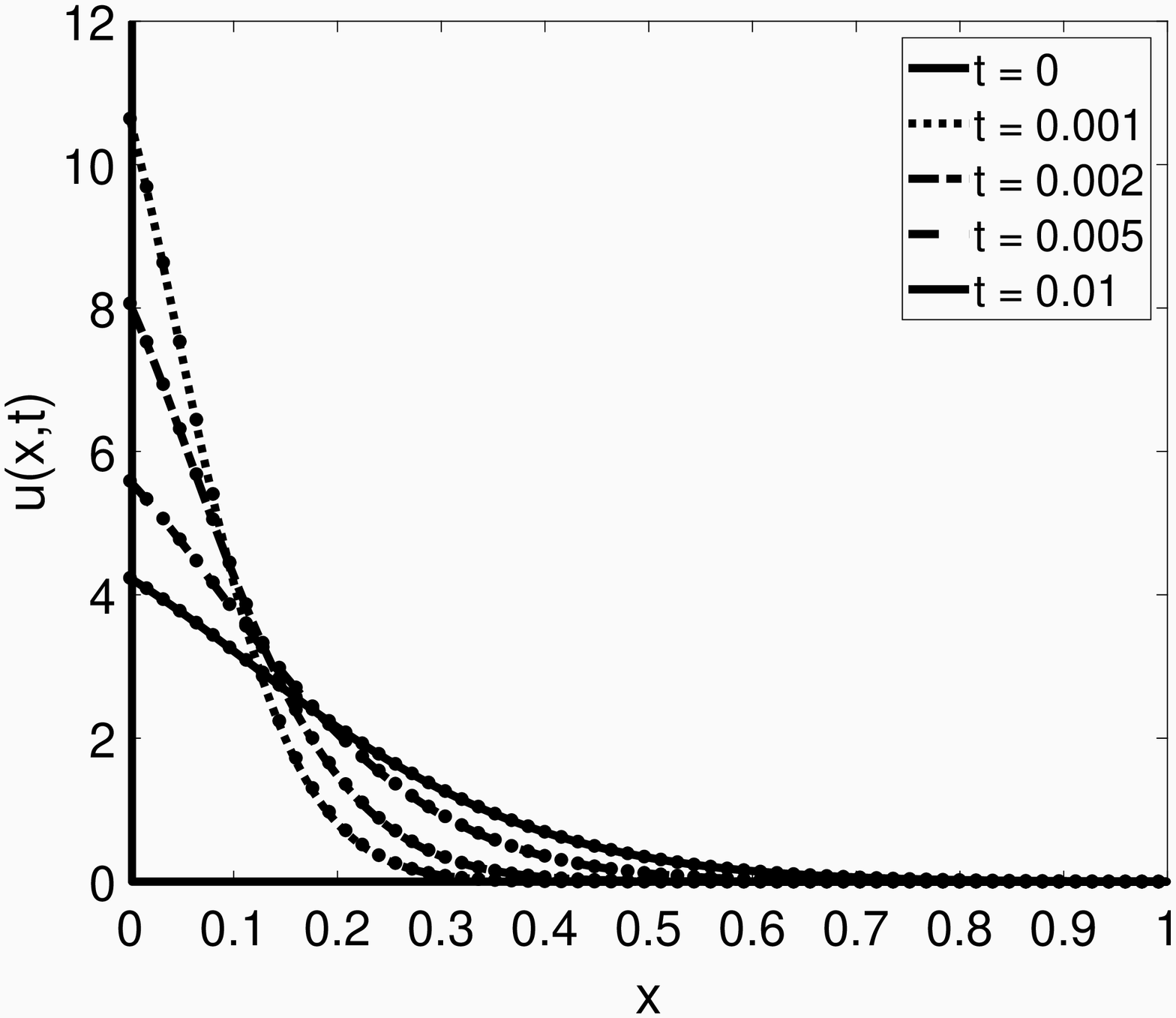}}
\vfil
\subfigure{\includegraphics[width=3in]{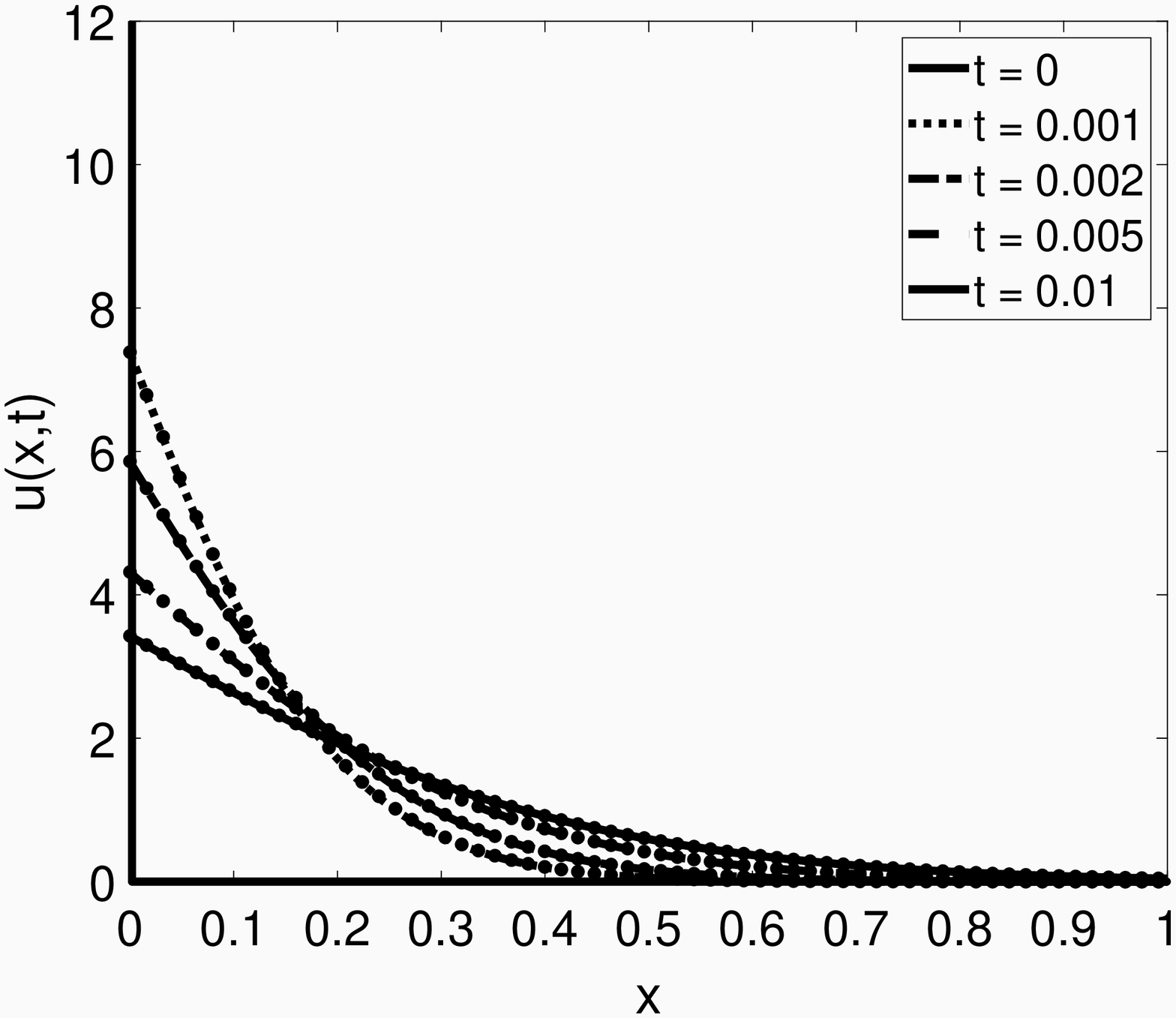}}
\caption{Numerical solutions of \eqref{fde2} with a reflecting boundary condition evaluated at $t = $ 0, 0.001, 0.002, 0.005, and 0.01 (lines) for $\alpha = 2$ (top left), 2.5 (top right), and 3 (bottom) and dual solution \eqref{invstablesub} (circles).} 
\label{stablesubfig}
\end{figure}

 As noted in \cite[Section 5]{baeumer2009space}, Equation \eqref{invstablesub} is the solution of the space-fractional PDE \eqref{fde2} on the half-line $x>0$. To make the problem \eqref{fde2} well-posed on the half-line \cite[Theorem 2.3]{baeumer2016reflected}, it is necessary to impose a fractional reflecting boundary condition given by \eqref{nofluxbc} at $x=0$ (see Appendix).  We numerically solved the negatively skewed space-fractional equation \eqref{fde2} subject to the reflecting boundary condition \eqref{nofluxbc} at $x=0$ on the domain $[0, 3]$ and an impulse initial condition using an implicit Euler scheme with reflecting (Neumann) boundary conditions outlined in \cite{baeumer2017boundary} and \cite{kellybcs2018}.  Since $2 \leq \alpha \leq 3$ in these examples, a shift of $m=1$ was applied to the Gr\"{u}nwald discretization.  The simulation was stopped before the signal reached the right boundary in order to mimic an infinite domain.  A total of $n=1501$ grid-points and a time step of $\Delta t = 0.00001$ was utilized to ensure sufficient accuracy.  Figure \ref{stablesubfig} displays these numerical solutions of \eqref{fde2} evaluated at $t = $ 0, 0.001, 0.002, 0.005, and 0.01 for $\alpha = 2$, 2.5, and 3, while the analytical solution \eqref{invstablesub} is shown in circles.  For $\alpha =2$, the solution is a normal density, while for $\alpha =3$, the dual solution is given by the Airy function \eqref{airysol}.  For $\alpha = 2.5$, the solution was checked against a numerical inverse Fourier transform
\begin{equation}
h_{\gamma}(x,t) = \frac{\alpha}{2 \pi} \int_{-\infty}^{\infty} \exp \left( t (-ik)^{\alpha} \right) \exp(i k x) \, dk
\label{invtransnum}
\end{equation}
evaluated using adaptive quadrature.  There is excellent agreement between the inverse stable density $h_{\gamma}(x,t)$ and these numerical solutions.
\section{Subordinated Brownian Motion in Multiple Dimensions}
All of the above examples are limited to one spatial dimension.  In this section, we show that multi-dimensional Brownian motion subordinated to a vector of independent inverse stable subordinators, defined by \eqref{invsub} in each dimension, is governed by a vector space-fractional PDE.  This multi-dimensional subordinated Brownian motion model may be useful for modeling contaminant transport in anisotropic media (multiscaling anomalous subdiffusion) \cite{zhang2006random}, where the retardation rate differs along each coordinate axis.
\subsection{Inverse stable subordinator vector}

Let $(u,v) = (E_t^1, E_t^2)$ be a pair of independent, inverse stable subordinators with densities $h_{\gamma_1}(u,t)$ and $h_{\gamma_2}(v,t)$ with indices $1/3 \leq \gamma_1 , \gamma_2<1$.  Physically, the indices $\gamma_1$ and $\gamma_2$ code the retention (retardation) that particles experience due to heterogeneity.  By independence, the joint density of $(E_t^1, E_t^2)$ is given by
\begin{equation}
h(u,v,t) = h_{\gamma_1}(u,t) h_{\gamma_2}(v,t) .
\label{jointdensity}
\end{equation}
Since the FLT of each density is $s^{\gamma_1 -1}/(s^{\gamma_1} + ik_u)$ and $s^{\gamma_2 -1}/(s^{\gamma_2} + ik_v)$, respectively, the convolution theorem \cite{howell_fourier} yields
\begin{equation}
\overline{h} (k_u, k_v, s) = s^{\gamma_1 -1}/(s^{\gamma_1} + ik_u) * s^{\gamma_2 -1}/(s^{\gamma_2} + ik_v)
\label{jointdensityconv}
\end{equation}
where the convolution ``*" is with respect to $s$.  Since the FLT is not a simple algebraic expression, it is difficult to find a simple time-nonlocal governing equation for the joint-density \eqref{jointdensity}.  

Although the order of the time fractional derivative in one dimension determines the retardation factor, here there are two different retardation factors, and only one time variable.  Hence, a time-fractional operator does not have enough degrees of freedom to code for both retardation factors.  However, we may find a vector space-fractional equation.  By space-time duality, each factor in \eqref{jointdensity} satisfies a space-fractional PDE
\begin{subequations}
\begin{equation}
\frac{\partial}{\partial t} h_{\gamma_1}(u,t) = \frac{\partial^{\alpha_1}}{\partial (-u)^{\alpha_1} } h_{\gamma_1} (u,t) \quad\text{for $u>0$ and }
\label{h1sp}
\end{equation}
\begin{equation}
\frac{\partial}{\partial t} h_{\gamma_2}(v,t) = \frac{\partial^{\alpha_2}}{\partial (-v)^{\alpha_2} } h_{\gamma_2} (v,t) \quad\text{for $v>0$.}
\label{h2sp}
\end{equation}
\label{hsp}
\end{subequations}
where $\alpha_1 = 1/\gamma_1$ and $\alpha_2 = 1/\gamma_2$.  Apply a time-derivative to \eqref{jointdensity} and apply the chain rule and \eqref{hsp}, yielding
\begin{equation}
\frac{\partial}{\partial t} h(u,v,t) = \frac{\partial^{\alpha_1} }{\partial (-u)^{\alpha_1} } h(u,v,t) + \frac{\partial^{\alpha_2} }{\partial (-v)^{\alpha_2} } h(u,v,t) 
\label{hgov}
\end{equation}
for $u>0$ and $v>0$, which is the space-fractional governing equation of the process $(E_t^1, E_t^2)$.  Note that \eqref{hgov} also governs \emph{operator stable L\'{e}vy motion} \cite{meerschaert2012} with backward, independent jumps in both the $u$ and $v$ directions. 
\subsection{Application to 2D Independent Brownian Motion}
Next, we consider a pair of independent Brownian motions subordinated (time-changed) by a pair of independent inverse stable subordinators.  Anisotropic super-diffusion may be modeled with the multi-dimensional fractional advection dispersion equation (FADE) \cite{meerschaert1999multidimensional}; however, we are not aware of any FPDE that models sub-diffusion in anisotropic media where the retardation factor in each coordinate is different.  In this section, we write the density of this 2D process, and determine the corresponding governing equation.

Let $x = B^{1}(u)$ and $y = B^{2}(v)$ be independent Brownian motions with densities $p(x,u)$ and $p(y,v)$, respectively.  Let $(u,v) = (E_t^1, E_t^2)$ be a pair of independent, inverse stable subordinators with densities $h_{\gamma_1}(u,t)$ and $h_{\gamma_2}(v,t)$ with indices $0 \leq \gamma_1 , \gamma_2<1$, respectively.  By a conditioning argument, we can write the joint density $q(x,y,t)$ of $\left( x , y \right) = \left( B^{1} \left( E_t^1 \right) , B^{2} \left( E_t^2 \right) \right)$ as
\begin{align*}
q(x,y,t) =& \left( \int_0^{\infty} p(x,u) h_{\gamma_1}(u,t) \, du \right)  \left( \int_0^{\infty} p(y,v) h_{\gamma_2}(v,t) \, dv \right) \\
         =&  \int_0^{\infty} \int_0^{\infty}  p(x,u)  p(y,v) h(u,v,t) \, du \, dv .
\end{align*}
The variables $u$ and $v$ are the temporal scaling of $B^{1}(u)$ and $B^{2}(v)$.  Hence, $q(x,y,t)$ is characterized by two time-scales. 
Using a partial fraction expansion, we may evaluate the subordination integrals above in closed form.  For example, 
\begin{equation}
\int_0^{\infty} p(x,u) h_{\gamma} (u,t) \,du = \frac{1}{2} h_{\gamma/2} \left( |x|,t \right) 
\label{eval1}
\end{equation}
and similarly for the $v$ integral.  Alternatively, one may use the composition formulas in Mainardi et al.\cite[Section 5]{mainardi2001fundamental} to derive \eqref{eval1}.  Applying \eqref{eval1} yields
\begin{equation}
q(x,y,t) = \frac{1}{4} h_{\gamma_1 / 2} \left(|x|,t \right) h_{\gamma_2 / 2} \left(|y|,t \right). 
\label{subbrownsol}
\end{equation}
Hence, the density $q(x,y,t)$ is symmetric about the $x$ and $y$ axes, but is not radially symmetric in general.  

Now let $2/3 \leq \gamma_1 , \gamma_2 \leq 1$ and $\alpha_1 = 2/ \gamma_1$ and $\alpha_2 = 2/ \gamma_2$.  For $x > 0$ and $y > 0$, space-time duality implies that the the density of each inverse stable subordinator satisfies 
\begin{subequations}
\begin{equation}
\frac{\partial}{\partial t} h_{\gamma_1 / 2} \left(x ,t \right) = \frac{\partial^{\alpha_1}}{\partial (-x)^{\alpha_1} } h_{\gamma_1 / 2} \left(x ,t \right) \quad\text{for $x>0$ and }
\label{dual1}
\end{equation}
\begin{equation}
\frac{\partial}{\partial t} h_{\gamma_2 / 2} \left(y ,t \right) = \frac{\partial^{\alpha_2}}{\partial (-y)^{\alpha_2} } h_{\gamma_2 / 2} \left(y ,t \right) \quad\text{for $y>0$.}
\label{dual2}
\end{equation}
\end{subequations}
Apply the product rule to \eqref{subbrownsol}, yielding
\begin{align*}
\frac{\partial}{\partial t} q(x,y,t) =& \frac{1}{4} \frac{\partial h_{\gamma_1 / 2} (x,t)}{\partial t} h_{\gamma_2 / 2} (y,t) + \frac{1}{4} h_{\gamma_1 / 2} (x,t) \frac{\partial h_{\gamma_2 / 2} (y,t)}{\partial t} \\
 =& \frac{1}{4} \frac{\partial^{\alpha_1}}{\partial (-x)^{\alpha_1}} h_{\gamma_1 / 2} \left(x ,t \right)  h_{\gamma_2 / 2} (y,t)  + \frac{1}{4} h_{\gamma_1 / 2} (x,t) \frac{\partial^{\alpha_2}}{\partial (-y)^{\alpha_2} } h_{\gamma_2 / 2} \left(y ,t \right) \\
 =&  \frac{\partial^{\alpha_1}}{\partial (-x)^{\alpha_1} } q(x,y,t) +  \frac{\partial^{\alpha_2}}{\partial (-y)^{\alpha_2} } q(x,y,t) 
\end{align*}
for $x>0$ and $y>0$.  Using the argument in Sec.~\ref{waveeqdual}, we see that $h_{\gamma_1/2}(x,t)$ satisfies \eqref{spacefrac2b} for $x<0$.  By the same token, $h_{\gamma_2/2}(y,t)$ satisfies a similar system of space-fractional equations, yielding the two-dimensional governing equation
\begin{equation}
\frac{\partial}{\partial t} q(x,y,t) = A^{\alpha_1}_x q(x,y,t) + A^{\alpha_2}_y q(x,y,t) ,
\label{subbrowngov}
\end{equation}
where $A_x^{\alpha}$ is defined by \eqref{Aoperator}.  The governing equation \eqref{subbrowngov} is the two-dimensional generalization of \eqref{ucomp}.  Since $2/3 \leq \gamma_1 , \gamma_2 \leq 1$, it follows that $2 \leq \alpha_1, \alpha_2 \leq 3$.  We conclude that the governing equation of $\left( B^{1} \left( E^{1}_t \right), B^{2} \left( E^{2}_t \right)  \right)$ is the space-fractional PDE \eqref{subbrowngov} utilizing both negative (right) and positive (left) RL fractional derivatives with orders greater than or equal to two.  This is another example of sub-diffusion modeled with a space-fractional PDE.  Generalization of \eqref{subbrowngov} to $n$-dimensional Brownian motion time-changed by $n$ independent inverse stable subordinators $\left( E^{1}_{\gamma_1} , \cdots, E^{n}_{\gamma_n} \right)$ is straightforward.
 
\begin{figure}
\subfigure{\includegraphics[width=3in]{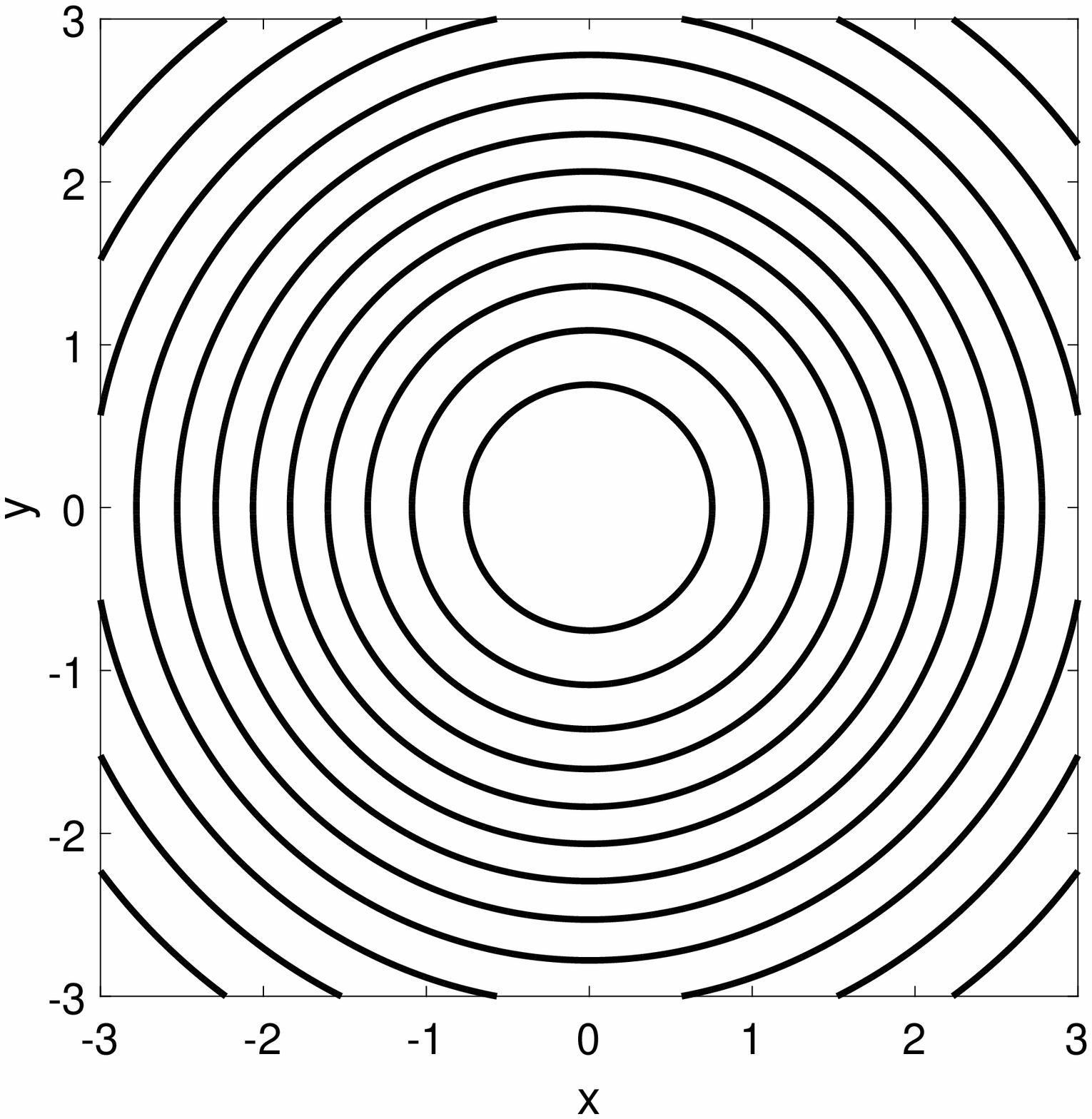}}
\hfil
\subfigure{\includegraphics[width=3in]{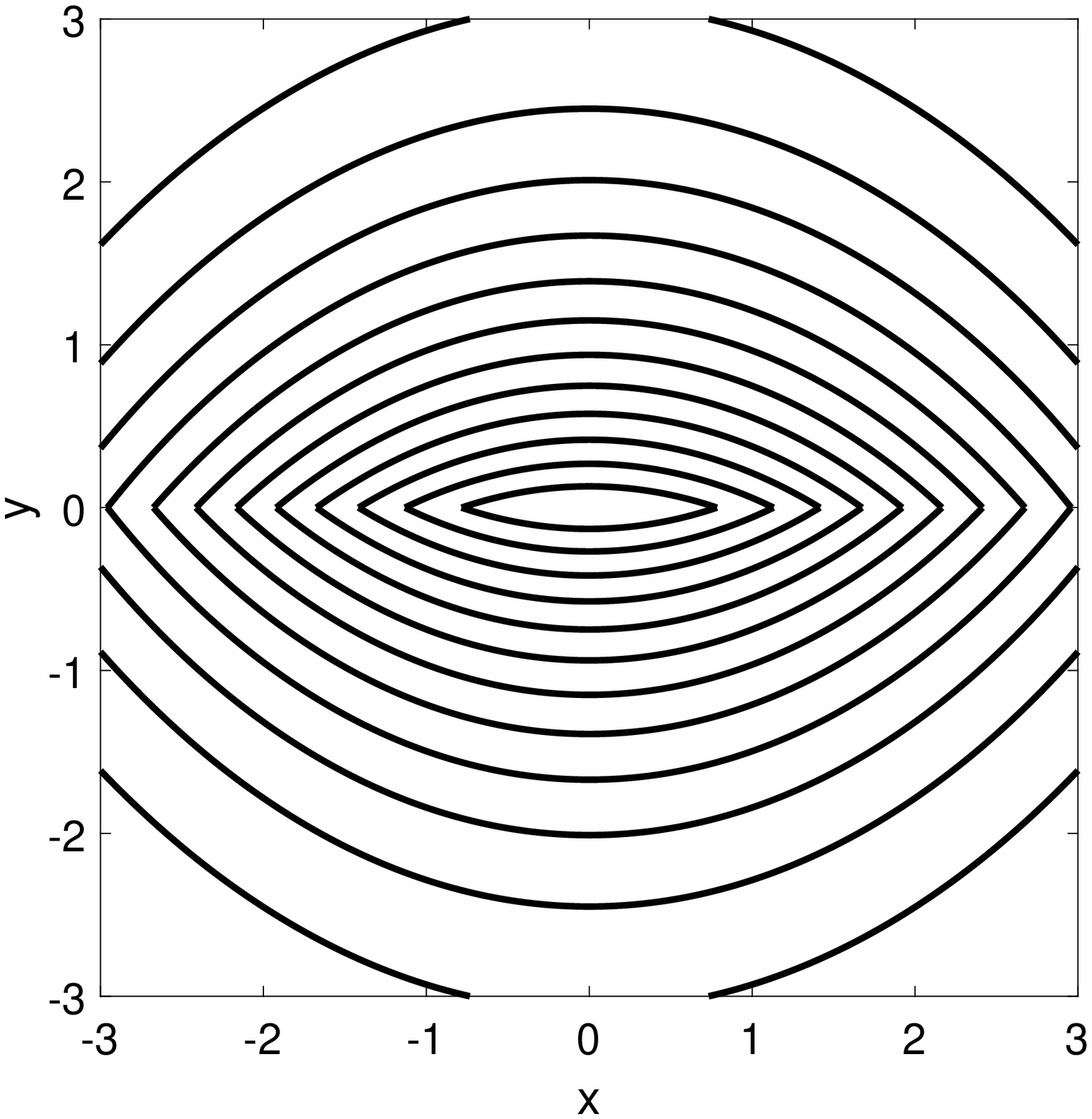}}
\vfil
\subfigure{\includegraphics[width=3in]{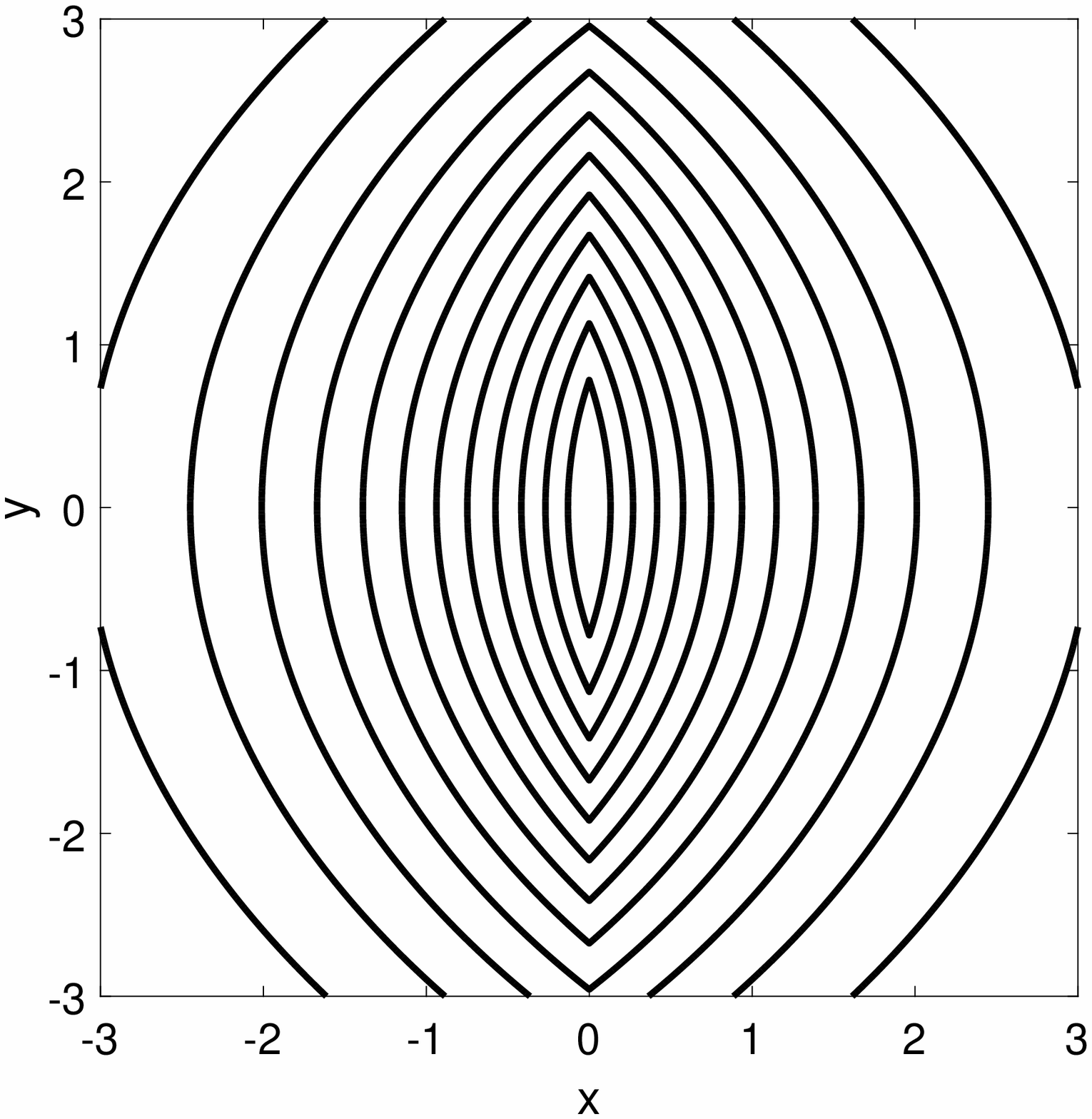}}
\hfil
\subfigure{\includegraphics[width=3in]{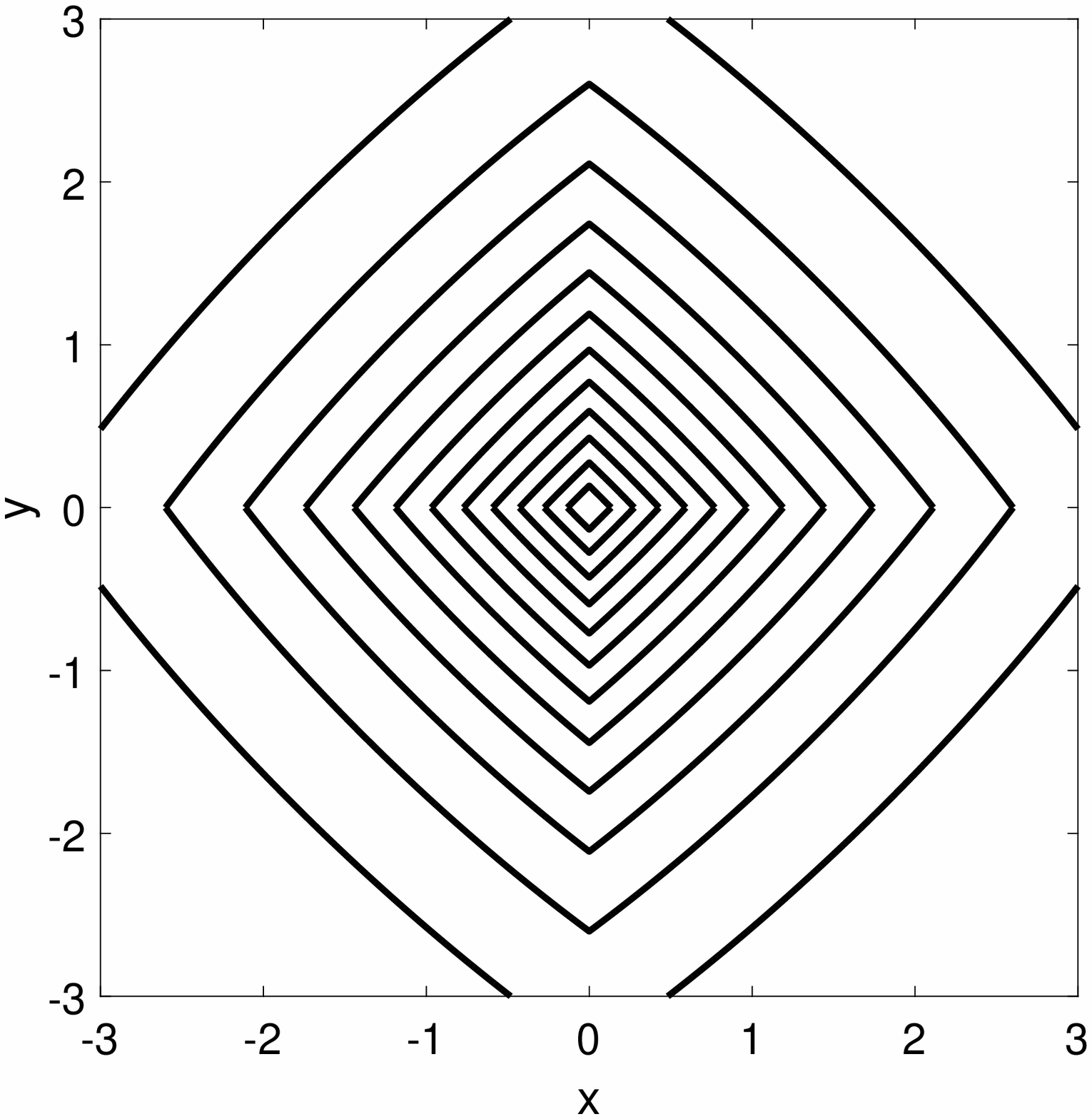}}
\caption{Contour plots of the solution \eqref{subbrownsol} to the fractional diffusion equation \eqref{subbrowngov} for delayed Brownian motion in two dimensions, with delay factors $\gamma_1 = 1$ and $\gamma_2 = 1$ (top left), $\gamma_1 = 1$ and $\gamma_2 = 0.5$ (top right), $\gamma_1 = 1$ and $\gamma_2 = 0.5$ (bottom left), and $\gamma_1 = 0.5$ and $\gamma_2 = 0.5$ (bottom right).}

\label{multibrownfig}
\end{figure}

Figure \ref{multibrownfig} displays contour plots of the joint density \eqref{subbrownsol} of $\left( B^1(E_t^1), B^2(E_t^2) \right)$ for Brownian motion $\gamma_1 = \gamma_2 =1$ (top left), sub-diffusion in the $x$ dimension and Brownian motion in the $y$ dimension $\gamma_1 = 0.5$ and $\gamma_2 = 1$ (top right), Brownian motion in the $x$ dimension and sub-diffusion in the $y$ dimension $\gamma_1 = 1$ and $\gamma_2 = 0.5$ (bottom left), and sub-diffusion along both axes $\gamma_1 = \gamma_2 = 0.5$ (bottom right).  Except for the top left panel (Brownian motion), these densities do not have radial symmetry, including the bottom right panel, where the inverse stable indices are the same in both directions.  In the case of sub-diffusion along both axes (bottom right), the density is not differentiable along the lines $x = 0$ and $y=0$, which follows from \cite[Proposition 6.1]{alrawashdeh2017applications}. 
%
%
\section{Tempered Duality}
Tempered fractional time derivatives impose an exponential cutoff to power-law waiting times \cite{alrawashdeh2017applications, meerschaert2008tempered}, while tempered fractional space derivatives cool power-law jumps in space \cite{baeumer2010tempered, cartea2007}.  Tempered fractional diffusion equations transition from anomalous to Fickian transport \cite{meerschaert2008tempered}.  This transition is governed by the spatial tempering rate $\lambda>0$ or the temporal tempering rate $\mu > 0$, which is typically small relative to the characteristic spatial or temporal scales, respectively.  For tempered space-fractional diffusion, the cross-over time (relaxation time) from anomalous to Fickian transport is proportional to $\lambda^{-\alpha}$, while for tempered time-fractional diffusion, the cross-over time is proportional to $\mu^{-1/\gamma}$ \cite{cartea2007}.  The tempering parameter also increases the effective diffusivity, which is given by Equation (31) in \cite{cartea2007}.  An alternative approach for modeling the transition from anomalous short time behavior to Fickian long term behavior are persistent random walks \cite{sadjadi2015}, where a self-propulsion mechanism competes with random fluctuations.

In this section, we apply space-time duality to connect tempered space-fractional and tempered time-fractional diffusion equations.  A negative Riemann-Liouville tempered fractional derivative of order $\alpha$ may be defined via
\begin{equation}
\frac{\partial^{\alpha, \lambda}}{\partial (-x)^{\alpha, \lambda}} f(x) = e^{\lambda x} \frac{\partial^{\alpha}}{\partial (-x)^{\alpha}} \left[ e^{-\lambda x} f(x) \right] - \lambda^{\alpha} f(x)
\label{negtemprl}
\end{equation}
where $\partial^{\alpha} / \partial (-x)^{\alpha}$ is the negative RL fractional derivative given by \eqref{negrl}.  The negatively-skewed tempered space-fractional diffusion equation is written using non-dimensionalized units as
\begin{equation}
\frac{\partial}{\partial t} u(x,t) = \frac{\partial^{\alpha, \lambda}}{\partial (-x)^{\alpha, \lambda}} u(x,t). 
\label{tsfrac}
\end{equation}
The second term in \eqref{negtemprl} is needed to ensure that solutions to \eqref{tsfrac} are proportional to a PDF (mass-conserving).  For $1 < \alpha \leq 2$, solutions to \eqref{tsfrac} with an impulse initial condition are given by \cite{baeumer2010tempered}
\begin{equation}
u(x,t) = e^{\lambda x} p(x,t) e^{-t \lambda^{\alpha}}
\label{usol11}
\end{equation}
where $p(x,t)$ is a negatively-skewed $\alpha$-stable density that solves \eqref{fde2}.  By space-time duality, $p(x,t)$ also satisfies 
\begin{equation}
\partial_t^{\gamma} p(x,t) = -\partial_x  p(x,t) 
\label{dualtemp}
\end{equation}
for $x>0$, where $\gamma = 1/ \alpha$ and the left hand side is the Caputo derivative or order $\gamma$.  Solving \eqref{usol11} for $p(x,t)$, inserting into \eqref{dualtemp}, and applying the product rule yields
\begin{equation*}
e^{-t \lambda^{\alpha}} \partial_t^{\gamma} \left[ e^{t \lambda^{\alpha}} u(x,t) \right] - \lambda u(x,t) = -\frac{\partial}{\partial x} u(x,t) .
\end{equation*}
Letting $\mu = \lambda^{\alpha}$, we see $u(x,t)$ solves the equivalent tempered time-fractional PDE
\begin{equation}
\partial_t^{\gamma, \mu} u(x,t) = -\frac{\partial}{\partial x} u(x,t)
\label{dualtemp2}
\end{equation}
where 
\begin{equation}
\partial_t^{\gamma, \mu} f(t) = e^{-\mu t} \partial_t^{\gamma} \left[ e^{t \mu} f(t) \right] - \mu^{\gamma} f(t) .
\label{temperedcaputo}
\end{equation}
Hence, the tempered time-fractional equation \eqref{dualtemp2} has the same solution as the tempered space-fractional equation\eqref{tsfrac}, where the tempering rates are related by $\mu = \lambda^{\alpha}$.  From a stochastic point of view, \eqref{tsfrac} governs tempered spectrally negative L\'{e}vy motion conditioned to stay positive with negative jumps, while \eqref{dualtemp2} governs power-law waiting times with an exponential cutoff.  By the equivalence between \eqref{dualtemp2} and \eqref{tsfrac}, backward jumps with power-law index $\alpha$ and tempering parameter $\lambda$ have the same governing equation as waiting times with power-law index $1/\alpha$ and tempering parameter $\lambda^{\alpha}$.  
%
%
\section{Conclusions}
This paper extends space-time duality to fractional diffusion for orders $1<\alpha  \leq 3$.  An equivalence with a time-fractional PDE is established using a Fourier-Laplace transform argument.  Since the equivalent time-fractional PDE governs the long-term limit of a power-law waiting time process, space-fractional diffusion equations with $2 <\alpha \leq 3$ gain a stochastic interpretation.  Using space-time duality, we show that the time-fractional diffusion-wave equation is a equivalent to a system of space-fractional diffusion equations.  Then we show that multi-dimensional Brownian motion subordinated to an independent inverse stable subordinator in each dimension is governed by a vector space-fractional PDE.  Finally, we extend the space-time duality to tempered fractional models for transient anomalous diffusion.
\begin{acknowledgments}
Kelly was partially supported by ARO MURI grant W911NF-15-1-0562 and USA National Science Foundation grant EAR-1344280.  Meerschaert was partially supported by ARO MURI grant W911NF-15-1-0562 and USA National Science Foundation grants DMS-1462156 and EAR-1344280.   Insightful discussion with Medhi Samimee (Department of Mechanical Engineering, Michigan State University) and Harish Sankaranarayanan (Department of Statistics and Probability, Michigan State University) are gratefully acknowledged.  We thank John Nolan (Department of Mathematics and Statistics, American University, Washington, DC ) for graciously providing the Stable toolbox ({\texttt www.RobustAnalysis.com}).
\end{acknowledgments}
\appendix*
\section{Reflecting Boundary Condition}
We demonstrate that \eqref{fde2} restricted to the half-line $x>0$ is equivalent to a boundary-value problem with a \emph{reflecting boundary condition} at $x=0$ \cite{baeumer2017diffeqs, baeumer2017boundary}.  Observe that since $h_{\gamma} (x,t)$ is a PDF with support on the half-line $x>0$, the total mass $\int_0^{\infty} u(x,t) \, dx$ on the half-line is a constant $\gamma$ for all times $t$.  Write \eqref{fde2} in a conservation form
\begin{equation}
\frac{\partial}{\partial t} u(x,t) = -C \frac{\partial}{\partial x} q(x,t)
\label{fde2cons}
\end{equation} 
where $q(x,t)$ is the fractional flux constitutive equation 
\begin{equation}
q(x,t) = C \frac{\partial^{\alpha -1}}{\partial (-x)^{\alpha -1}} u(x,t) ,
\label{flulxfunc}
\end{equation} 
which has been proposed for super-diffusion ($\alpha <2$) by Paradisi et al. \cite{paradisi2001fractional} and Schumer et al. \cite{schumer2001eulerian} and for hyperdiffusion ($\alpha > 2$) by Wei \cite{wei1999generalized} and Hu et al. \cite{hu2013high}.  Due to the factor of $(-1)^n$ in \eqref{negrl}, the derivative of the $(\alpha -1)$ negative RL derivative is $-\frac{\partial^{\alpha}}{\partial (-x)^{\alpha}}$.  Assuming that $u(x,t)$ is bounded for $t>0$, we have
\begin{equation*}
\frac{\partial}{\partial t} \int_0^{\infty} u(x,t) \, dx = - \int_0^{\infty} \frac{\partial}{\partial x} q(x,t) \, dx = q(0,t) ,
\end{equation*}
where the flux is assumed to be zero at infinity.  Mass conservation on $x>0$ yields the no-flux (or reflecting) boundary condition
\begin{equation}
\frac{\partial^{\alpha -1}}{\partial (-x)^{\alpha -1}} u(0,t) = 0 ,
\label{nofluxbc}
\end{equation}
which were studied by Baeumer et al. \cite{baeumer2017diffeqs}. 
\bibliographystyle{unsrt}

%

\end{document}